\documentstyle[12pt]{article}
\title{Vacuum energy and Universe in special relativity}
\author{G.E. Volovik\\
Low Temperature Laboratory,
Helsinki University of Technology\\
P.O.Box 2200, FIN-02015 HUT, Finland\\
and\\
L.D. Landau Institute for Theoretical Physics,
   Moscow\\
}
\begin{document}
\maketitle
\begin{abstract}
{ The   problem of cosmological constant and vacuum energy is
usually thought of as the subject of general relativity. However,
the vacuum energy is important for the Universe even in the
absence of gravity, i.e. in the case when the Newton constant
$G$ is exactly zero, $G=0$.  We discuss the response of the
vacuum energy to the perturbations of the quantum vacuum in
special relativity, and find that as in general relativity the
vacuum energy density is on the order of the energy density of
matter.  In general
relativity, the dependence of the
vacuum energy on the equation of state of matter
does not contain
$G$, and thus is valid in the limit
$G\rightarrow 0$. However, the result obtained for the
vacuum energy in the world without gravity, i.e. when $G=0$
exactly, is different.
   }
\end{abstract}

The problem of the vacuum energy appears to be more general than
the cosmological constant problem which arises in general
relativity \cite{Weinberg2}. Earlier we discussed the vacuum energy
and its relation to the cosmological constant considering
general relativity as an effective theory \cite{Book,Phenomenology}.
We
found that the energy of the equilibrium vacuum state at zero
temperature is
zero due to the general thermodynamic Gibbs-Duhem relation applied to
the
vacuum as a medium.  The non-zero value of the vacuum energy, or more
exactly
the gravitating part of the vacuum energy, comes from perturbations
of the
vacuum state.  In typical situations the perturbations of the vacuum
are
caused by the gravitating matter, and thus the induced vacuum energy
density
must be on order of the energy density of matter, which results in the
cosmological constant consistent with observations
\cite{RiessPerlmutter}.

Now we
extend the discussion of the vacuum energy to the case of special
relativity,
i.e.  to the Universe without gravity. Though in the world without
gravity
the cosmological constant is absent, the vacuum energy still plays an
important role in the structure of the Universe.  We find how the
vacuum
energy responds to matter in  special relativity, and how this allows
us to
stabilize the static special-relativity Universe filled with matter
having
arbitary equation of state.

The cosmological term in the
action for the general relativity is
\begin{equation}
   S_\Lambda=
-\int d^4x\sqrt{-g}\rho^{\rm vac}~.
\label{EinsteinAction3}
\end{equation}
The corresponding stress-energy
tensor of the vacuum is obtained by variation of the action over
the metric tensor $g^{\mu\nu}$
\begin{equation}
   T^{\rm vac}_{\mu\nu}=\rho^{\rm vac}u_\mu u_\nu+
P^{\rm vac}(u_\mu u_\nu-g_{\mu\nu})
=\rho^{\rm vac} g_{\mu\nu}
   ~.
\label{VacuumEM}
\end{equation}
Here $\rho^{\rm vac}$ is
the vacuum energy density and  $u^\mu$    the
4-velocity of the vacuum. Since the equation of state for the
vacuum is
\begin{equation}
P^{\rm vac}=-\rho^{\rm vac}
   ~,
\label{VacuumEM1}
\end{equation}
the energy-momentum
tensor does not depend on  the 4-velocity  $u^\mu$, and thus
is the same in any coordinate system.

The Eq.(\ref{VacuumEM})  with equation of state (\ref{VacuumEM1})
are valid for the vacuum in special relativity too, i.e. in the
absence of the dynamical field $g^{\mu\nu}$. These equations are
obtained by the conventional procedure used in quantum field
theory, when one introduces the fictituous field, such as the
fictituous gauge fields or fictituous metric, and calculate the
response of the vacuum to these fields.  Moreover, the equation
of state
$P^{\rm vac}=-\rho^{\rm vac}$ is even more general, since it is
valid even in the non-relativistic theories, where the
quantum vacuum is played by the ground state of the
quantum condensed matter. This equation of state comes
from the general thermodynamic Gibbs-Duhem relation applied to
the homogeneous ground state of a condensed matter (see. e.g.
\cite{Phenomenology}).

Let us consider the Universe in special relativity
(i.e. in the absence of gravity, $G=0$), which is filled with
non-gravitating homogeneous matter -- the perfect cosmic
fluid, and discuss how the vacuum responds to the matter and
stabilizes this
Universe. The energy-momentum tensor of matter is
\begin{equation}
   T^{\rm M}_{\mu\nu}=\rho^{\rm M}v_\mu v_\nu+
P^{\rm M}(v_\mu v_\nu-g_{\mu\nu})
   ~,
\label{MatterEM}
\end{equation}
where  $v^\mu$    the
4-velocity of matter, and $\rho^{\rm M}$ and
$P^{\rm M}$ are energy density and pressure of matter in the
comoving frame.

In the general coordinate frame, the energy and momentum density
of matter are
\begin{equation}
   \tilde \rho^{\rm M}= {\rho^{\rm M} +{v^2\over
c^2}P^{\rm M}\over 1- {v^2\over
c^2}} ~~,~~{\bf p}^{\rm M}=  {{\bf v}\over c^2}~{\rho^{\rm M}
+P^{\rm M} \over 1- {v^2\over
c^2}}~.
\label{MatterEnergyMomentum}
\end{equation}
The obvious consequence of Eq.(\ref{MatterEnergyMomentum}) is
that the energy and the momentum of matter do not satisfy the
relativistic relation between the energy and momentum
\begin{equation}
    {\bf P} = E {{\bf v}\over c^2}~.
\label{LorentzEnergyMomentum}
\end{equation}
The reason for that is related to the external forces acting on
matter, which
violate the Lorentz invariance, since they establish the preferred
reference
frame in which these forces are isotropic.  These forces are
presented in
Eq.(\ref{MatterEnergyMomentum}) through the pressure $P^{\rm
M}$ of matter, which is supported by the external pressure
(see Sec. 14 of Ref. \cite{Einstein1907}). If the Universe is
completely
isolated from the "environment", the external pressure is absent,
$P^{\rm M}=P_{\rm external}=0$, and the Lorentz-invariant
equation (\ref{LorentzEnergyMomentum}) is restored.  But the
typical matter considered in cosmology, such as the relativistic
plasma, does not exist at zero pressure as an equilibrium state,
except for the extreme limit case of the cold matter.  Thus,
within the special relativity the Universe must be either
empty, or contain such a matter which can exist in equilibrium at zero
pressure (the matter in a cold liquid state, for example).

The vacuum gives an alternative scenario for the equilibrium static
Universe
with matter to exist in special relativity.  The equilibrium state is
achieved when the pressure of the cosmological matter is compensated
by
the partial pressure of the vacuum, so that the external pressure
becomes
zero:
\begin{equation}
P_{\rm external}=P^{\rm vac}+P^{\rm M}=0 ~,
\label{VacuumEM2} \end{equation} and the Universe (matter + vacuum)
can be in
equilibrium without external environment.  For this equilibrium
Universe the
Eq.(\ref{LorentzEnergyMomentum}) for the energy and momentum of the
whole
Universe is also restored.

Using equation of state (\ref{VacuumEM1}) and the equilibrium
condition (\ref{VacuumEM2}) one obtains  the density of the vacuum
energy
induced by matter with pressure $P^{\rm M}$ in the equilibrium
Universe:
\begin{equation}
\rho^{\rm vac}=P^{\rm M}
   ~.
\label{VacuumEM3}
\end{equation}
Since the vacuum momentum is zero
\begin{equation}
{\bf p}^{\rm vac}=0~,
\label{VacuumMomentum}
\end{equation}
the total energy density and momentum density of the system
(matter + vacuum) become
\begin{equation}
\rho_{\rm total}= \tilde \rho^{\rm M}+\rho^{\rm vac}=
{\rho^{\rm M} +\rho^{\rm vac}\over 1- {v^2\over
c^2}} ~~,~~{\bf p}_{\rm total}= {\bf p}^{\rm M}+{\bf p}^{\rm
vac}=  {{\bf v}\over c^2}~{\rho^{\rm M} +\rho^{\rm vac} \over 1-
{v^2\over c^2}}~.
\label{TotalEnergyMomentum}
\end{equation}
They satisfy the relativistic  equation
(\ref{LorentzEnergyMomentum}), and
the corresponding density of the rest mass  of the system is
\begin{equation}
\rho_{\rm total} = {m_{\rm rest}c^2\over \sqrt{1- {v^2\over
c^2}}} ~~,~~m_{\rm rest}=
{\rho^{\rm M} +\rho^{\rm vac}\over c^2\sqrt{1- {v^2\over
c^2}}} ~.
\label{TotalRestMass}
\end{equation}
The extra factor  $\sqrt{1-  v^2/c^2}$ in the denominator of
the rest mass is cancelled by the relativistic transformation of
the volume: the volume
$V$ in the frame of the measurement and the volume
$V_{\rm comoving}$ in the comoving frame are related as
$dV=dV_{\rm comoving}\sqrt{1-  v^2/c^2}$, so that the total rest
energy of the system is
\begin{equation}
M_{\rm rest}=\int dV
{\rho^{\rm M} +\rho^{\rm vac}\over \sqrt{1- {v^2\over
c^2}}}= V_0
\left(\rho^{\rm M} +\rho^{\rm vac}\right)  ~.
\label{TotalRestEnergy}
\end{equation}
Thus the whole world represents the relativistic object whose
rest mass is the sum of the rest energies of the matter and
quantum vacuum.
The energy density of the quantum vacuum induced by the nongravitating
($G=0$) matter is completely determined by the equilibrium
condition (\ref{VacuumEM2}) and
equation of state for matter $P^{\rm M}=w^{\rm M}\rho^{\rm M}$:
\begin{equation}
\rho^{\rm vac}_{G=0}=w^{\rm M}\rho^{\rm M} ~.
\label{VacuumEnergyG0}
\end{equation}

We can compare the vacuum energy (\ref{VacuumEnergyG0}) in the
Universe in special relativity, i.e. when $G=0$, with the vacuum
energy in Universes in general relativity, i.e. when $G\neq 0$. For
the Einstein static closed Universe
\cite{EinsteinCosmCon,Phenomenology} the vacuum energy induced by the
gravitating matter is
  \begin{equation} \rho^{\rm vac}_{\rm Einstein}={1\over
2}\left(1+3w^{\rm M}\right)\rho^{\rm M} ~,
\label{VacuumEnergyEinstenU}
\end{equation}
and for the G\"odel steady-state rotating Universe
\cite{Goedel,Phenomenology} it is:
\begin{equation}
\rho^{\rm vac}_{\rm Goedel}=-{1\over 2}\left(1-w^{\rm
M}\right)\rho^{\rm M}
   ~.
\label{VacuumEnergyGoedelU}
\end{equation}
In all three Universes, the density of the vacuum energy
induced by matter is proportional to the energy density of
matter. Equations (\ref{VacuumEnergyEinstenU}) and
(\ref{VacuumEnergyGoedelU}) for the worlds with gravity do not
depend on the Newton constant
$G$, and thus are valid in the limit  $G\rightarrow 0$. But they
do not coincide with Eq.(\ref{VacuumEnergyG0}) for
the world without gravity, i.e. when
$G$ is exactly zero. While in special relativity the
vacuum response to matter is determined by the condition of zero
external pressure, $P_{\rm external}=0$, in the case of the
gravitating matter the condition of gravineutrality is added
\cite{Phenomenology}. The  pressure and energy of
the gravitational field contributes to both conditions even
in the limit $G\rightarrow 0$. These  contributions come from
the space curvature in the Einstein Universe and from the local
rotational
metric in the G\"odel Universe.

As distinct from the static gravitating Universes which experience
different
types of the gravitational instabilities, in special relativity the
Universe
with matter is stable, if the quantum vacuum itself is the stable
object. In
the latter case, small perturbations of the vacuum state, caused by
cold or
hot matter, do not destabilize the system. This fact is well known
for the
condensed matter analogs of the Universe --  quantum liquids, where
the
role of the quantum vacuum is played by the superfluid condensate,
and the
role of the relativistic matter is played by the "relativistic"
quasiparticles, with $c$ being the maximum attainable speed of the
low-energy
quasiparticles.  Examples are provided by the Bose-superfluid
$^4$He and Fermi-superfluid
$^3$He-A.
Both quantum liquids are stable at $T=0$ and $P=0$, and their
stability is not violated by the massless "relativistic"
quasiparticles which
appear at $T\neq 0$ forming the analog of matter in these toy
Universes. For
both liquids the equation (\ref{VacuumEnergyG0}) is valid with the
"relativistic" equation of state $w^{\rm M}=1/3$, if the liquids
are isolated from the environment
\cite{Book}.  Though in both
quantum liquids there are the low-frequency collective modes
corresponding to
the dynamics of the effective metric, this effective gravity does not
obey
the general covariance and is not Newtonean at large distances. As a
result
the effective gravity in these liquids does not modify the
special-relativistic equation
(\ref{VacuumEnergyG0}).

For such
condensed matter systems, the relativistic equations
(\ref{TotalEnergyMomentum}) obtained for the energy and
momentum of a Universe in special relativity are also
applicable, but with one reservation.  As distinct from its
special
relativity counterpart, the quantum vacuum (condensate) in condensed
matter
does not obey the effective Lorentz invariance obeyed by the
excitations of the condensate -- quasiparticles.  In particular, the
momentum
density of the quantum condensate is non-zero. As distinct from
Eq.(\ref{VacuumMomentum}), the superfluid condensate moving with the
so-called superfluid velocity ${\bf u}_{\rm s}$ carries the
momentum density
${\bf p}^{\rm vac}=mn{\bf u}_{\rm s}$, where $m$ is the mass of
particles
comprizing the condensate (atoms of liquid), and $n$ is their number
density.
As a result, for quantum liquids the relativistic equations are valid
only in
the reference frame moving with the condensate. The full
correspondence
between the quantum vacuum and superfluid condensate could occur only
for
such hypothetical condensates whose "atoms" are massless, $m=0$.
However,
the difference between the relativistic quantum vacuum and
non-relativistic quantum condensate does not change the
conclusion that the vacuum response stabilizes
the non-gravitating Universe.

I thank A.A. Starobinsky for fruitful discussions. This work was
supported by
ESF COSLAB Programme and by the Russian Foundations for
Fundamental Research.

\end{document}